%% file: main.tex
\documentclass[reprint,aps,prl,amsmath,longbibliography,superscriptaddress, floatfix]{revtex4-1}
\usepackage{color}
 \usepackage{ulem}
\usepackage{graphicx}
\usepackage{dcolumn}
\usepackage{bm}
\usepackage{float}
\usepackage[binary-units=true]{siunitx}
\usepackage{braket}
\usepackage{ mathrsfs }
\usepackage{enumitem}
\usepackage[dvipsnames]{xcolor}
\usepackage{bbold}
\usepackage{url}
\usepackage[nohyperlinks]{acronym}
\usepackage{threeparttable}
\include{config-pdflatex}

\usepackage[colorlinks=true,
  urlcolor=blue,
  linkcolor=blue,
  pdfborderstyle=,
  citecolor=blue]{hyperref}

\DeclareSIUnit{\liter}{l}

\setlist[description]{style=unboxed, leftmargin=7pt, itemsep=0.3ex, itemindent=\dimexpr\labelwidth+\labelsep\relax}

\begin{document}
\setlength{\parskip}{0.2cm}
\title{\headerfont \Large The HDvent Emergency Ventilator System}
\author{David Grimshandl}
    \affiliation{Physics Institute, Heidelberg University, 69120 Heidelberg, Germany}
\author{Manuel Gerken}
    \affiliation{Physics Institute, Heidelberg University, 69120 Heidelberg, Germany}
\author{Eleonora Lippi}
    \affiliation{Physics Institute, Heidelberg University, 69120 Heidelberg, Germany}
\author{Binh Tran}
    \affiliation{Physics Institute, Heidelberg University, 69120 Heidelberg, Germany}
\author{Saba Zia Hassan}
    \affiliation{Physics Institute, Heidelberg University, 69120 Heidelberg, Germany}
\author{Jan Hendrik Becher}
    \affiliation{Physics Institute, Heidelberg University, 69120 Heidelberg, Germany}
\author{Selim Jochim}
    \affiliation{Physics Institute, Heidelberg University, 69120 Heidelberg, Germany}
\author{Matthias Weidem\"{u}ller}
    \affiliation{Physics Institute, Heidelberg University, 69120 Heidelberg, Germany}
\author{Gunnar F\"{o}hner}
    \affiliation{Physics Institute, Heidelberg University, 69120 Heidelberg, Germany}
\author{Steffen Brucker}
    \affiliation{Physics Institute, Heidelberg University, 69120 Heidelberg, Germany}
\author{Frank Schumacher}
    \affiliation{Physics Institute, Heidelberg University, 69120 Heidelberg, Germany}
\author{Wolfgang Beldermann}
    \affiliation{Physics Institute, Heidelberg University, 69120 Heidelberg, Germany}
\author{Venelin Angelov}
    \affiliation{Physics Institute, Heidelberg University, 69120 Heidelberg, Germany}
\author{Stefan Hetzel}
    \affiliation{Physics Institute, Heidelberg University, 69120 Heidelberg, Germany}
\author{Stefan Hummel}
    \affiliation{Physics Institute, Heidelberg University, 69120 Heidelberg, Germany}
\author{Simon Muley}
    \affiliation{Physics Institute, Heidelberg University, 69120 Heidelberg, Germany}
\author{Bernd Windelband}
    \affiliation{Physics Institute, Heidelberg University, 69120 Heidelberg, Germany}
\author{Christoph Eisner}
    \affiliation{Heidelberg University Hospital, 69120 Heidelberg, Germany}
\author{Marco Zugaj}
    \affiliation{Heidelberg University Hospital, 69120 Heidelberg, Germany}
\author{Philipp Bayer}
    \affiliation{Rommelag iLabs GmbH, 76133 Karlsruhe, Germany}
\author{Dario Ernst}
    \affiliation{Rommelag iLabs GmbH, 76133 Karlsruhe, Germany}
\author{Marcel Gehrlein}
    \affiliation{Rommelag iLabs GmbH, 76133 Karlsruhe, Germany}
\author{Helmut Jacob}
    \affiliation{Rommelag iLabs GmbH, 76133 Karlsruhe, Germany}
\author{Jens Pfeifle}
    \affiliation{Rommelag iLabs GmbH, 76133 Karlsruhe, Germany}
\author{Andreas Treskatsch}
    \affiliation{Rommelag iLabs GmbH, 76133 Karlsruhe, Germany}
\author{Martin Ulmschneider}
    \affiliation{Rommelag iLabs GmbH, 76133 Karlsruhe, Germany}
\author{Philipp M. Preiss}    \email{preiss@physi.uni-heidelberg.de}
    \affiliation{Physics Institute, Heidelberg University, 69120 Heidelberg, Germany}

\date{
\today}

\begin{abstract}
\noindent The pandemic caused by \ac{sars-cov2} has affected countries all across the world, heavily burdening the medical infrastructure with the growing number of patients affected by the coronavirus disease (\acs{cv19}). With ventilators in limited supply, this public health emergency highlights the need for safe, fast, reliable, and economical alternatives to high-end commercial devices and has prompted the development of easy-to-use and mass-producible ventilators. Here, we detail the design of the \textit{HDvent Emergency Ventilator System}. The device performs ventilation through mechanical compression of manual resuscitators and includes control electronics, flow and pressure sensors, and an external data visualization and monitoring unit. We demonstrate its suitability for open loop, pressure- and volume-controlled ventilation. The system has not undergone clinical testing and has not been approved for use as a medical device.
 The project documentation needed to reproduce the prototype is freely available and will contribute to the development of open source ventilation systems.

\end{abstract}
\maketitle
\section{Introduction}
The \ac{who} declared a worldwide pandemic in March 2020, due to the rapid and, to a large extent, uncontrolled propagation of \ac{sars-cov2} or \acs{cv19} \cite{fisher2020}. As of \today, the virus has infected over 75 million people with about 1.7 million reported deaths.

For many \acs{cv19} patients, suffering from severe cases of respiratory failure, life-saving treatment can be provided by oxygenation and mechanical ventilation \cite{maclaren_fisher_brodie_2020,Ranney2020}. While previous pandemics have already caused concerns about shortage of mechanical ventilators \cite{smetanin2009, mcneil2006}, the \acs{cv19} pandemic shows once more that the demand for mechanical ventilators in emergency situations can rapidly exceed the local availability in hospitals \cite{rubinson2010,NPR,BRANSON2020,beitler2020}, possibly leading to preventable deaths and triage of patients. Shortages of ventilators can occur in developing countries, but also in nations with fully developed medical infrastructure. There, mechanical ventilators are often highly complex devices that can only be provided by a few specialized suppliers and at long lead times, making a dynamic response to increased demand difficult.

The present pandemic therefore highlights the need of an additional class of ventilators beyond the high-end devices used in typical ICU settings: Openly available designs for cost-effective, easily mass-producible ventilators are needed for rapid emergency responses. Such devices should cover the basic medical requirements for mechanical ventilation, but may not need the full set of features of a high-end ventilator \cite{BRANSON2020}. Intended for the treatment of infected, but less critical cases, such ventilators may free up resources for the treatment of critical cases and also mitigate the burden on the medical infrastructure.

In view of the ongoing coronavirus pandemic and with the urge to concretely contribute to the eventuality of a local emergency situation, a group of technicians, physicists, anesthesiologists and IT professionals came together to develop a working prototype of such a mechanical ventilator, the \textit{HDvent Emergency Ventilator System}, which we describe in this article.
\begin{figure*}
	\centering
	\includegraphics[width = 0.99\textwidth]{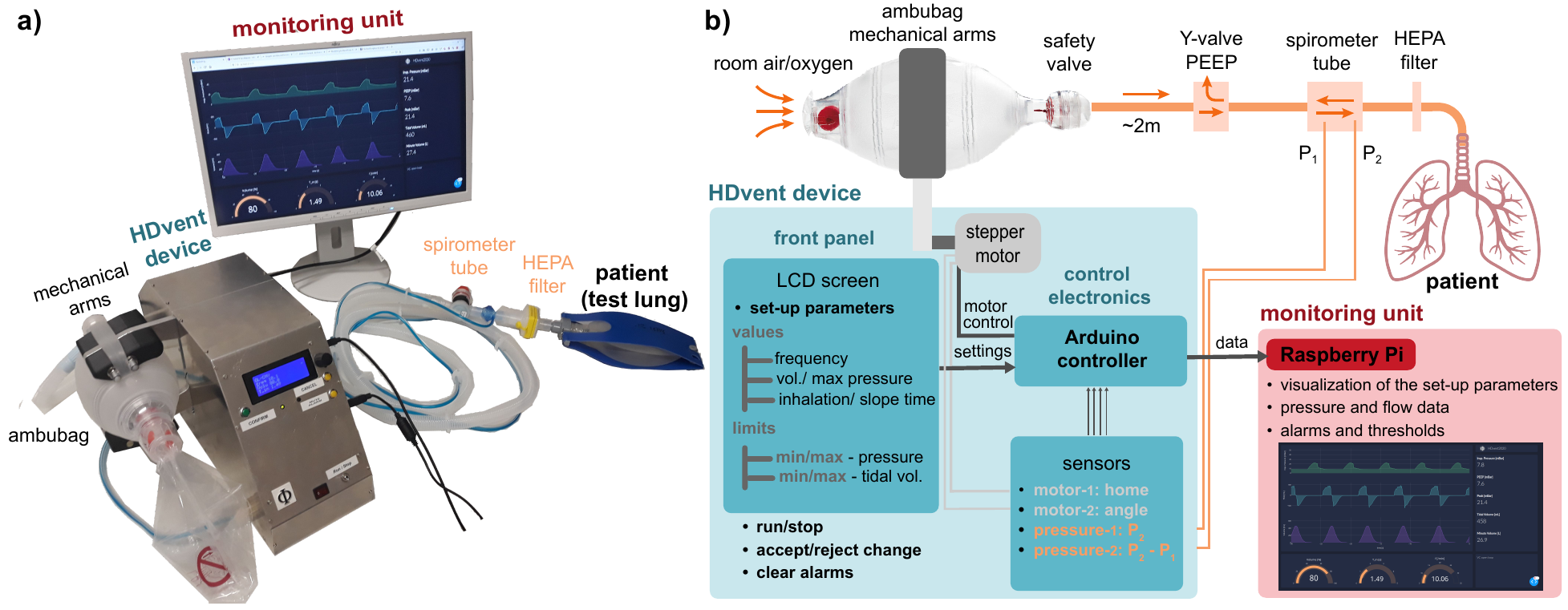}
	\caption{Overview of the HDvent system. a) Photograph showing the device housing a stepper motor as well as control electronics and sensors on a single PC board mounted to the front panel. Air is supplied to the patient via compression of the externally mountable \acl{ambu} and standard hoses and valves. Ventilation parameters and alarm thresholds are accessible via the user interface on the front panel. b) Schematic of the HDvent architecture. All components directly in contact with the patient are single-use (orange). An optional monitoring unit (red) can be attached to the main device (blue) to display ventilation parameters, alarm thresholds and time traces.}
	\label{fig:block_diagram}
\end{figure*}
\subsection{The HDvent system}
The key design goals of the HDvent system are: 
\begin{description}
    \item[Simplicity] The ventilator should be compatible with mass production and require only modest machining capabilities. 
    \item[Durability] The device should be suited for prolonged mechanical ventilation over two weeks or more, which may require several hundred thousand ventilation cycles. 
    \item[Ease of Operation] The operation of the device should be intuitive for medical staff or even untrained personnel. Potentially fatal user errors should be prevented by design.
    \item[Cost] For mass-scale production or deployment in developing countries, the cost of the device should not exceed several hundred USD, a fraction of the typical price of several ten thousand dollars of commercial ventilators \cite{BRANSON2020}.
\end{description}
To meet the above criteria for an emergency ventilator system, several approaches have been proposed \cite{review_open_ventilators,Harris2020,Mora2020}. One strategy is to directly supply the patient with compressed air and/or oxygen, which can yield compact devices, but requires access to compressed air supplies and can be susceptible to dangerous pressure spikes \cite{cern_ventilator,milano_ventilator,nasa_ventilator,portugese_ventilator,corovent,Raymond2020}. 

Our chosen approach is based on manual resuscitators, e.g. \textit{Ambubags}. Such resuscitators consist of an elastomer balloon of roughly \SI{1.5}{\liter} volume. The bag can be compressed manually to deliver tidal volumes of up to \SI{1}{\liter} of room air or an oxygen-enriched mixture to intubated adults. Ambubags are for single-patient use, cheap (can be less than ten USD), and available in large quantities in most major healthcare facilities. We follow this approach motivated by a similar use of ambubags by teams around the world \cite{oxvent_ventilator, rice_ventilator,open_ventilator,marburg_ventilator, MADVent}. Our design is inspired particularly by the work of the MIT e-vent team and aims to improve several aspects of the design \cite{EVent-2020}.
\section{Methods}
Figure \ref{fig:block_diagram} shows an overview of the components of the HDvent system (see caption). The different parts of the system are described in more detail below.\\
\subsection{Air management}
We work with the Dr\"{a}ger Oxylog 2000 Ventstar\textsuperscript{\textregistered} breathing circuit and the Ambu SPUR II system, but other resuscitators can be used. The bag takes in room air or optionally oxygen-enriched room air via a reservoir. It is followed by a \SI{40}{\milli\bar} over-pressure safety valve and 2 metres of flexible hose. Near the patient a membrane Y-Valve separates in- from exhaled air, which is critical to reduce dead volume in the breathing circuit. The open end of the Y-Valve is piloted with a mechanical valve from the \acl{ambu} system to set the \ac{peep}. 
The flow rate of in- and exhaled air is monitored with a spirometry tube. The pressure is picked up near the patient mouth piece using two thin (\SI{2}{\milli\metre} ID) hoses connected to the HDvent ($\text{P}_1$ and $\text{P}_2$ in Fig.\ref{fig:block_diagram}b)). We measure the pressure relative to atmosphere and the differential pressure $P_\textrm{2}-P_\textrm{1}$ across a thin mesh in the spirometry tube, which is directly proportional to the flow.  A HEPA filter between the valve and the patient ensures that the in- and exhaled air is free of virus material and particulate contaminants. All parts in contact with the patient or breathable air are available commercially and approved for clinical use.
\subsection{Mechanical design}
\begin{figure}
	\centering
	\includegraphics[width = 0.45\textwidth]{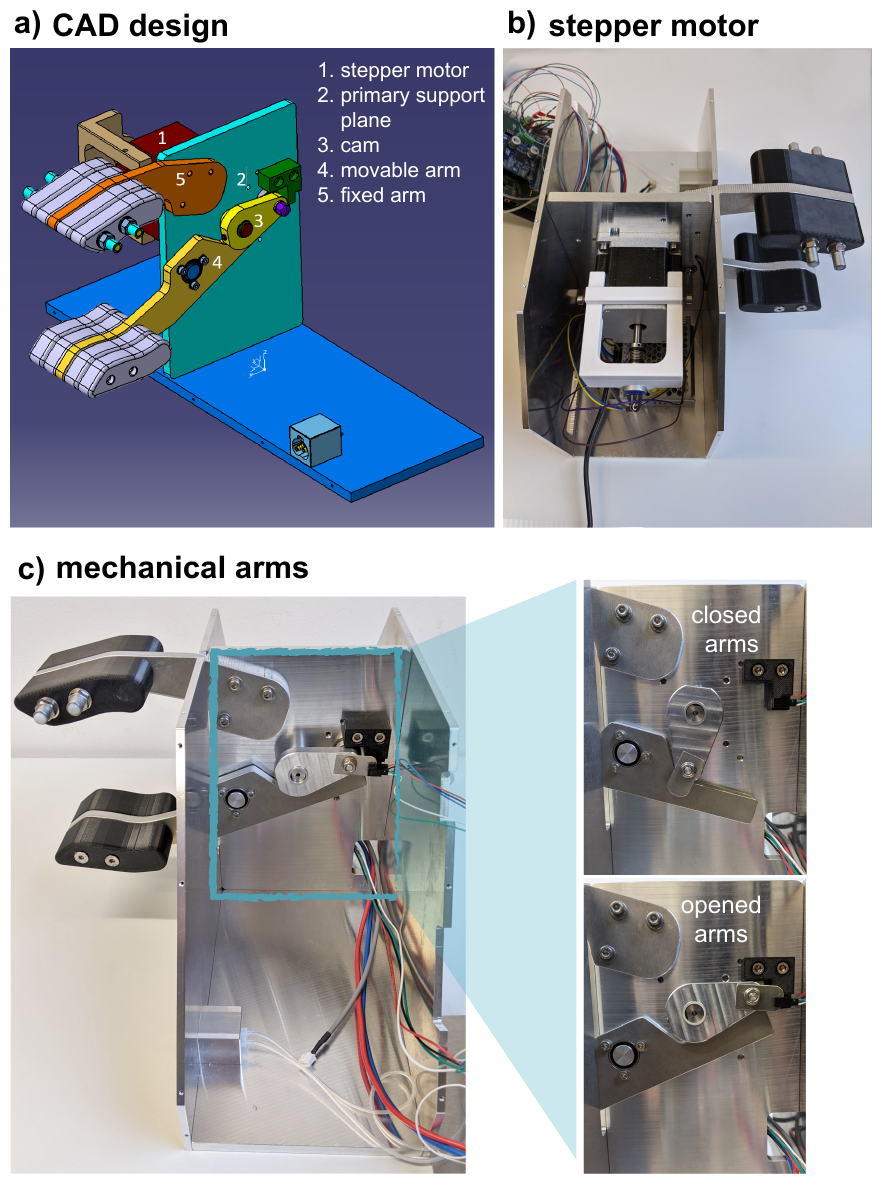}
	\caption{Mechanical design of the ventilator. a) CAD model of the main components. A stepper motor (1) mounted to the primary support plate (2) drives a cam (3) and the movable arm (4) against the fixed arm (5). b) Photograph showing the mounting of the stepper motor and c) the mechanical arms in closed and open positions.}
	\label{fig: mechanics}
\end{figure}
Fig.\ref{fig: mechanics} shows a CAD drawing and photographs of the main mechanical parts of the HDvent device. The compression system consists of two mechanical arms and is driven by a NEMA 34 stepper motor rated at \SI{4}{\newton\meter} holding torque. Different from the MIT design, where both arms are actuated by gear wheels, we fix one arm and allow the other one to rotate on a ball bearing. A second bearing on the end of a crank mounted on the motor shaft runs on the flat upper surface of the pivoting arm, forcing it to rotate. Apart from the off-the-shelf ball bearings the only part subject to friction is the running surface of the actuated arm. This simple design allows to manufacture both the arms and the mounting and base plate from the same \SI{10}{\milli\metre} aluminium stock, for example with a water jet. The device is completed with a sheet metal enclosure which provides electrical grounding and protection from pinch hazards. 
\subsection{Electronics Design and Sensors}
\begin{figure*}
	\centering
	\includegraphics[width = 0.99\textwidth]{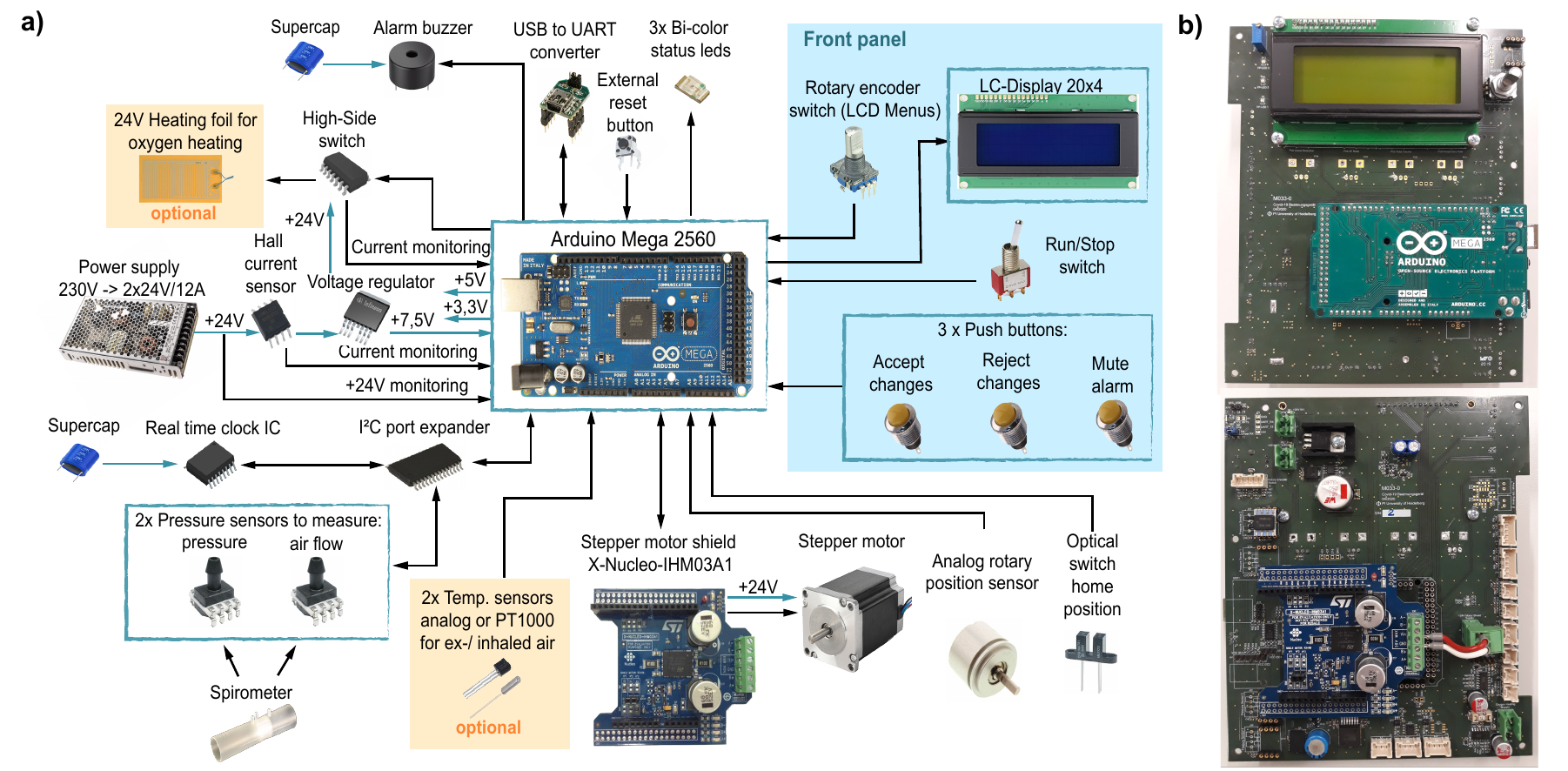}
	\caption{Electronics design. a) Block diagram showing the power electronics, sensors, and user controls. The device is controlled by an Arduino Mega microcontroller. Connections for a \SI{24}{\volt} battery or for a second, redundant power supply as well as optional features, such as temperature monitoring and heating of the air/oxygen supply, are already implemented on the board. b) Photographs showing the front and back of the custom-made board, which is mounted directly to the front panel of the HDvent unit.}
	\label{fig: electronics}
\end{figure*}

The HDvent is controlled by an ATMega2560 microcontroller equipped on an Arduino Mega 2560 board, which handles motor control, user input, sensor readings, power management and data flow. The microcontroller, peripheral electronics, and sensors are mounted on a  PC board. The motor is controlled via the powerSTEP01 stepper controller chip on an evaluation board, which supports microstepping up to $1/128$, features a sensorless stall detection and is controlled via an SPI interface. Fig. \ref{fig: electronics} shows an overview of the implemented features and photographs of the custom-made board.

For pressure and flow measurements, we use two board-mounted differential pressure sensors from the Honeywell HSC series. We selected a sensor with range \SI{\pm100}{\milli\bar} for the inspiratory pressure measurement and a second sensor with a finer range of \SI{\pm16}{\milli\bar} for the spirometry. Several sensors monitor the proper operation of the device: An optical switch defines the home position of the stepper and an analog resistive rotation sensor coupled to the motor shaft allows absolute measurements of the stepper position. A Hall current sensor monitors the built-in \SI{24}{\volt} modular power supply.
\label{sec: electronics}
\subsection{Microcontroller Software}
The controller software is written in C++ using the standard Arduino library for all I/O operations. It uses a state machine architecture to control the ventilation cycle and to provide the user interface. One execution of the main program loop takes about \SI{15}{ms}. This constitutes also the sampling rate for pressure and flow measurements. Similarly the cycle time determines the jitter of time based triggers in the ventilation cycle and imposes an upper limit on the bandwidth of the PID algorithm (for closed loop modes).

Commercial ventilators support a range of different ventilation modes which can be categorized by the following attributes \cite{nomenclature_draeger}:
\begin{description}[before={\renewcommand\makelabel[1]{\bfseries ##1}}]
\item[Subset of trigger methods] triggering the inspiration and expiration phase
\item[Feedback variable and setpoint profile] used to calculate the motor speed, either running the PID algorithm or applying the setpoint curve in open loop control.
\item[Set of control parameters] matching the selected mode. They are displayed and can be edited via the user interface.
\end{description}
In our implementation new ventilation modes are defined in terms of those attributes. This high level of abstraction allows to easily add ventilation modes without changing the core controller software. So far three ventilation modes have been implemented (using the nomenclature proposed in \cite{nomenclature_draeger}):
\begin{description}
\item[\acs{ol-cmv} \acl{ol-cmv}] begin of inspiration is timer triggered, length of inspiration is fixed, the stepper is moved at constant speed. 
\item[\acs{pc-cmv} \acl{pc-cmv}] pressure is controlled following a trapezoidal setpoint ramp, begin of inspiration is timer triggered, length of inspiration is fixed. 
\item[\acs{vc-cmv} \acl{vc-cmv}] flow is controlled following a trapezoidal setpoint ramp, begin of inspiration is timer triggered, length of inspiration is fixed. 
\end{description}
Special attention was also paid to the safety and reliability of the controller unit: A watchdog timer limits the execution time and automatically reboots the controller after program freezes. After reboot the controller resumes the  respiration without further user input, even after a hard shut off or a power outage, using the non-volatile \acs{eeprom} of the microcontroller to store ventilation parameters. During power-off an on-board supercapacitor powers a buzzer for an acoustic alarm signal. The ventilation control relies crucially on the position of the compressing arm. We use triple modular redundancy to determine the absolute position of the stepper motor: Values from the analog rotary position sensor, step counting by the stepper driver and an optical switch are processed by a triple voting logic. 
\subsection{HDvent User Interface}
\newcommand{\ui}[1]{\texttt{#1}}
The device is operated using controls and a LCD on the slanted front panel of the casing. A rotary encoder is used to scroll through a list of user-editable entries shown on the display. Menu items are (in this order):
\newcommand{\paramTin}{$T_\mathrm{in}$}
\newcommand{\paramRR}{$RR$}
\newcommand{\paramPinsp}{$P_{\mathrm{insp}}$}
\newcommand{\paramSlope}{$T_\mathrm{slope}$}
\begin{description}
    \item[ventilation mode] select from implemented modes: \acs{pc-cmv}, \acs{ol-cmv}, \acs{vc-cmv}
    \item[control parameters] multiple entries, depending on selected ventilation mode, e.g. for the \acs{pc-cmv} mode the operator can set the inspiration time, the respiratory rate and the slope and plateau value of the pressure ramp.
    \item[alarm thresholds] one entry for every diagnostic measurement eligible for alarms, e.g.\ minute volume or inspiratory pressure. The operator can set lower and upper alarm thresholds for every item.
\end{description}
\subsection{Monitoring Unit}
\begin{figure}
	\centering
	\includegraphics[width = 0.45\textwidth]{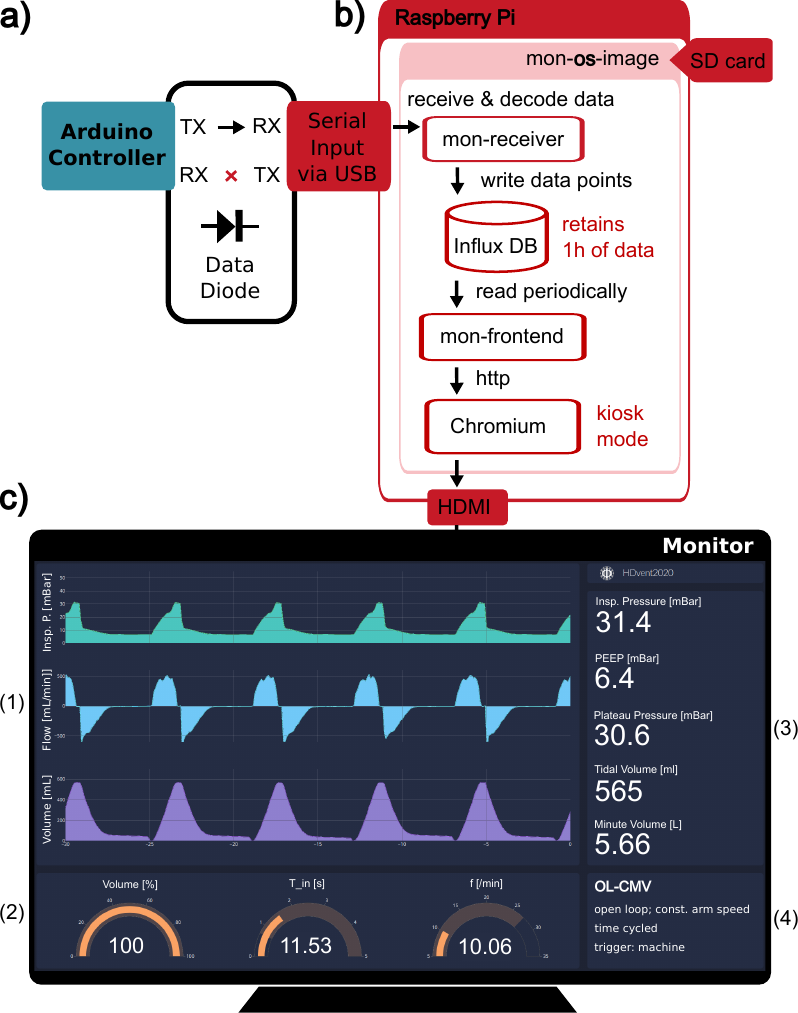}
	\caption{Monitoring unit hardware components and data flow. a) Data are transferred from the Arduino controller to the Raspberry Pi via a unidirectional serial link. b) On the Raspberry Pi, the data are processed, stored and a dashboard on an external display is generated, c). (1) time traces (\SI{30}{\second}) of inspiratory pressure, flow and volume, (2) ventilator controller settings, (3) numerical indicators for the remaining diagnostics, (4) current ventilation mode with short description}
	\label{fig: monitoring}
\end{figure}
The monitoring unit is an optional component to display information on the machine state and patient condition to HDvent operators. 
A \ac{pi} (Model 4B, Rev.1, \SI{4}{\giga\byte}) serves as the main controller. It is connected to the ventilator controller via a USB to \acs{uart} (\acl{uart}) bridge in a simplex transmission setup, establishing a unidirectional serial link that prevents any detrimental influence on ventilation control by system failure and incorrect or malicious operation of the monitoring unit (Fig.\,\ref{fig: monitoring}a)). 
\newcommand{\monComp}[1]{\texttt{#1}}
The software system is built on a customized \acl{pi} \ac{os} image \cite{github_HDvent}. Unnecessary components were removed from the \ac{os} image to reduce its size and attack surface in networked setups. On boot the \ac{os} automatically starts up all services and applications required to ingest incoming ventilator data and generate the dashboard view (Fig.\,\ref{fig: monitoring}b) and c)): The service \monComp{mon-receiver} checks incoming serial transmissions for errors and decodes the \acs{cobs} (\acl{cobs}) encoded serial stream into named data points. The decoded and timestamped data points are written to a time-series database (\monComp{InfluxDB})
that stores one hour of historical data.

The application \monComp{mon-frontend} periodically reads and aggregates data from the database, derives values such as minimum or maximum values and serves a web interface built with Dash. The interface is rendered on the \ac{pi} in the web browser Chromium,  running in kiosk mode to output to a standard computer monitor connected to the HDMI port in full screen.

The interface combines graphical representations of user configurable ventilator controller settings, time traces of inspiratory pressure, flow and volume and numerical indicators for the remaining diagnostics. Visual alarms display when preset thresholds are exceeded. The interface updates automatically once per second. It is read-only by design and cannot alter database entries or the ventilation controller settings.
\section{Results}

The full HDvent system was tested using a commercially available test lung (Dr\"{a}ger SelfTestLung\textsuperscript{TM}, $V_{max}$=\SI{1000}{\milli\liter}) which models resistance and compliance of an average adult. We calibrated the volume flow measurement to an estimated accuracy of \SI{\pm 2}{\percent} by forcing calibrated volumes of air through the spirometry tube.
\begin{figure}
	\centering
	\includegraphics[width = 0.48 \textwidth]{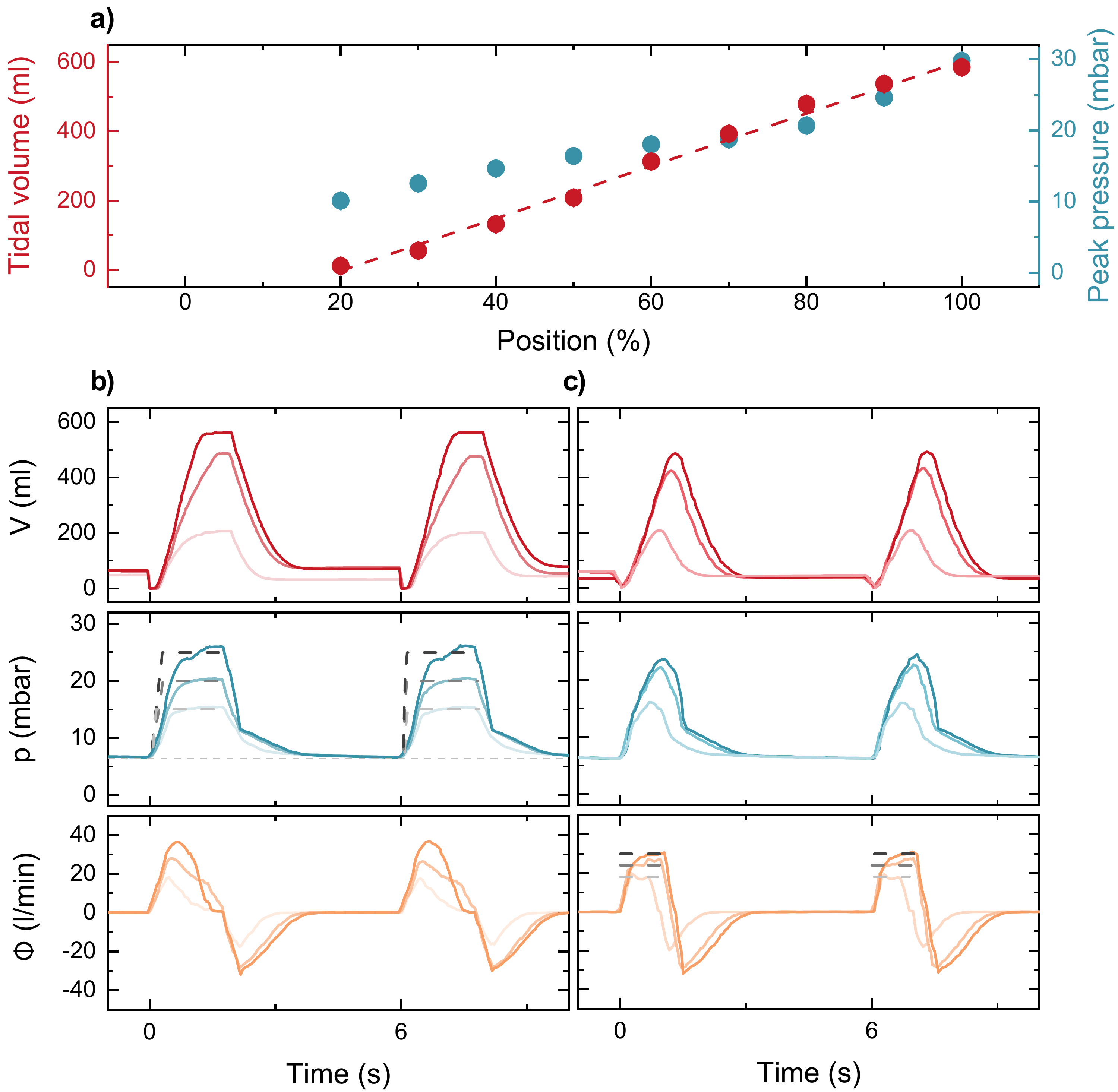}
	\caption{Ventilation performance. a) Measurement of tidal volume (red) and peak pressure (blue) over percentage of maximal displacement of the moving arm in \acs{ol-cmv} mode. The red line indicates a linear dependence of the tidal volume on the arm movement. The attained peak pressure reflects the compliance of the text lung. b) Volume $V$ (red), pressure $p$ (blue) and flow $\Phi$ (orange) over time in \acs{pc-cmv} mode for three different set pressures of \SI{25}{\milli\bar} (dark blue), \SI{20}{\milli\bar} (light blue) and \SI{15}{\milli\bar} (bright blue) shown as dashed lines. c) Ventilation in \acs{vc-cmv} mode for three different set flows of \SI{30}{\litre\per\minute} (dark orange), \SI{24}{\litre\per\minute} (light orange) and \SI{18}{\litre\per\minute} (bright orange) shown as dashed lines. For b) and c) a PEEP of \SI{7}{ \milli\bar} was set and BPM is 10.}
	\label{fig: data}
\end{figure}
We first measured the mechanical characteristics of the device in  \acs{ol-cmv} mode, i.e.\ at fixed stepper motor speeds. Fig.\ref{fig: data}\,a) shows the attained tidal volume and peak pressure versus the stepper motor angle (as a percentage of the maximum angle) for a PEEP of \SI{10}{\milli\bar}. The tidal volume follows the stepper angle amplitude in close to linear fashion. For the chosen combination of \acl{ambu}, test lung and PEEP, we obtain a maximal tidal volume of \SI{600}{\milli\liter}. We reach a max plateau pressure of \SI{40}{\milli\bar}, limited by the pressure relief valve of the \acl{ambu}. The measured \ac{peep} matches the pressure set on the mechanical \ac{peep} valve near the patient mouth piece.

The time traces in Fig.\ref{fig: data}\,b) show the ventilation characteristics in \acs{pc-cmv} mode. The measured pressure tracks the  setpoint (shown as the trapezoidal dashed line in the second panel) closely, indicating the proper performance of the feedback loop. Fig.\ref{fig: data}\,c) shows similar time traces for \acs{vc-cmv} mode. Here, the measured flow follows the set point indicated by dashed lines in the bottom panel. Table\,\ref{tab: parameters} shows the range of ventilation parameters accessible with the HDvent.

 \begin{table}[h]
 \begin{threeparttable}
 \def\arraystretch{1.1}
 \caption{Accessible range of common ventilation variables.}
 \begin{tabular}{p{2.5cm}ls[table-unit-alignment = left]} 
 
    respiratory rate & \numrange{10}{30}&BPM \\
    tidal volume & \numrange{0}{600}&\milli\liter \\
    minute volume & \numrange{0}{18}&\liter\per\minute\\
    peak pressure & \numrange{0}{40}&\milli\bar\tnote{a}\text{  }  \\
    \ac{peep} &  \numrange{0}{20}&\milli\bar\tnote{b}\text{  }  \\
\end{tabular}
\begin{tablenotes}\footnotesize 
\item[a] upper limit depends on the used overpressure valve
\item[b] range given by mechanical \ac{peep} valve
\vspace{0.1cm}
\end{tablenotes}
	\label{tab: parameters}
	\end{threeparttable}

	\end{table}

We performed a long-term stability measurement in OL-CMV. Over 3 days (43k cycles), the tidal volume and pressure profiles remained constant to \SI{\pm 0.5}{\percent}, while the device showed no signs of wear. The most limiting component in terms of longevity is the Ambubag, which we estimate should be replaced every 5 days for safety.

\section{Conclusions}
We have developed  a complete design for a simple and economical ventilator based on the \acl{ambu} system. The device is easy to manufacture and relies on off-the-shelf components and open source hard- and software. It was shown to perform volume- and pressure-controlled ventilation on a test lung within realistic parameter ranges. The architecture of the control software allows for the easy addition of other ventilation modes, e.g. with patient-triggered inspiration. The optional Pi based monitoring unit provides the operator with a graphical dash board familiar from commercial devices. Its automated build routine that forwards software changes directly to the OS image makes future extensions, like network based remote monitoring, particularly easy to implement and deploy.

The course of the \acs{cv19} pandemic beyond its first wave stretched the capabilities of the health care systems in developed countries locally, but fortunately did not require the widespread use of improvised emergency equipment such as the system presented here and elsewhere \cite{review_open_ventilators}. As such, our system has not yet been tested in a clinical setting and will not be submitted for approval as a medical device. Further testing could focus on prolonged stress tests, emergency protocols in case of component failures, spontaneous/assisted ventilation modes and ease of operation.

We hope that our design may inspire the development of low-cost ventilator solutions for pandemic emergencies and for regions with limited medical services. The full documentation needed to reproduce the prototype, including drawings, step files, and source code, is freely available \cite{github_HDvent}. We explicitly encourage interested developers to reach out for further details. 
\section{Key Findings}
\begin{itemize}
    \item We present an open-source design of a mechanical ventilator costing $\sim \SI{500}{\$}$ in materials.
    \item The device meets the clinical demands for the ventilation of patients with \acs{cv19}.
    \item The strengths of our approach are: (i) 
    a mechanical design with low part count and generous tolerances
    (ii)  a monitoring unit displaying all relevant data, and (iii) a flexible software architecture and user interface to accommodate different ventilation modes.
    \item We expect these advantages to contribute to the global efforts towards making open-source mechanical ventilators available for emergency use and for developing countries. 
\end{itemize}
\section{Acknowledgments} We would like to thank the members of the mechanical and electronic workshop, the construction and the purchasing department of the Physics Institute at Heidelberg University for their support. We are grateful to Würth Elektronik GmbH and WEDirekt for their generous support in realising the control electronics. 
\acrodef{ol-cmv}[OL-CMV]{Open loop continuous mandatory ventilation}
\acrodef{pc-cmv}[PC-CMV]{Pressure controlled continuous mandatory ventilation}
\acrodef{vc-cmv}[VC-CMV]{Volume controlled continuous mandatory ventilation}
\acrodef{who}[WHO]{World Health Organization}
\acrodef{ambu}{Ambubag}
\acrodef{peep}[PEEP]{positive end-expiratory pressure}
\acrodef{eeprom}[EEPROM]{Electrically Erasable Programmable Read-Only Memory}
\acrodef{cv19}[COVID-19]{covid-19}
\acrodef{sars-cov2}[SARS-CoV-2]{Severe Acute Respiratory Syndrome Coronavirus 2}
\acrodef{ol}[OL]{open loop mode}
\acrodef{pc}[PC]{pressure controlled mode}
\acrodef{vc}[VC]{volume controlled mode}
\acrodef{hdmi}[HDMI]{High Definition Multimedia Interface}
\acrodef{http}[HTTP]{Hypertext Tranfer Protocol}
\acrodef{os}[OS]{Operating System}
\acrodef{pi}[Pi]{Raspberry Pi}
\acrodef{uart}[UART]{Universal Asynchronous Receiver Transmitter}
\acrodef{cobs}[COBS]{Consistent Overhead Byte Stuffing}
\bibliography{hdvent_bibliography.bib}
\bibliographystyle{unsrt}

\end{document}

%% file: config-pdflatex.tex
\usepackage{microtype}

\usepackage{CrimsonPro}

\usepackage{FiraSans}

\newcommand{\headerfont}{\sffamily}

\usepackage{titlesec}

\titlespacing*{\section}
{0pt}{18pt plus 6.5pt minus 0.5pt}{7pt plus 0.3pt}
\titlespacing*{\subsection}
{0pt}{10pt plus 3.5pt minus 0.5pt}{5pt plus 0.3pt}

\titleformat{\section}%
  {\normalfont\headerfont\bfseries\Large}{}{10pt}{}
  
\titleformat{\subsection}%
  {\normalfont\headerfont\large}{}{10pt}{}
  

%% file: main.bbl
\begin{thebibliography}{10}

\bibitem{fisher2020}
Dale Fisher and David Heymann.
\newblock {Q\&A}: The novel coronavirus outbreak causing {COVID-19}.
\newblock {\em \href{https://doi.org/10.1186/s12916-020-01533-w}{BMC
  Medicine}}, 18:57, 2020.

\bibitem{maclaren_fisher_brodie_2020}
Graeme Maclaren, Dale Fisher, and Daniel Brodie.
\newblock {Preparing for the Most Critically Ill Patients With COVID-19}.
\newblock {\em \href{https://pubmed.ncbi.nlm.nih.gov/32074258/}{Jama}},
  323:1245, 2020.

\bibitem{Ranney2020}
Megan~L. Ranney, Valerie Griffeth, and Ashish~K. Jha.
\newblock {Critical Supply Shortages — The Need for Ventilators and Personal
  Protective Equipment during the Covid-19 Pandemic}.
\newblock {\em \href{https://doi.org/10.1056/NEJMp2006141}{New England Journal
  of Medicine}}, 382(18):e41, 2020.

\bibitem{smetanin2009}
Paul Smetanin, David Stiff, and et~al. Anand~Kumar.
\newblock {Potential Intensive Care unit Ventilator Demand/Capacity Mismatch
  due to Novel Swine-Origin H1N1 in Canada}.
\newblock {\em \href{https://doi.org/10.1155/2009/808209}{Canadian Journal of
  Infectious Diseases and Medical Microbiology}}, 20(4):e115--e123, 2009.

\bibitem{mcneil2006}
Donald~G. McNeil~Jr.
\newblock {Hospitals Short on Ventilators if Bird Flu Hits}, 2006.
\newblock Available:
  \url{https://www.nytimes.com/2006/03/12/us/hospitals-short-on-ventilators-if-bird-flu-hits.html}.

\bibitem{rubinson2010}
Lewis Rubinson, Frances Vaughn, Steve Nelson, and et~al. Giordano, Sam.
\newblock {Mechanical Ventilators in US Acute Care Hospitals}.
\newblock {\em
  \href{https://www.cambridge.org/core/journals/disaster-medicine-and-public-health-preparedness/article/mechanical-ventilators-in-us-acute-care-hospitals/F1FDBACA53531F2A150D6AD8E96F144D}{Disaster
  Med Public Health Prep.}}, 4:199–206, 2010.

\bibitem{NPR}
Patti Neighmond.
\newblock {As The Pandemic Spreads, Will There Be Enough Ventilators?}, 2020.
\newblock Available:
  \url{https://www.npr.org/sections/health-shots/2020/03/14/815675678/as-the-pandemic-spreads-will-there-be-enough-ventilators},
  [Accessed: Nov 2020].

\bibitem{BRANSON2020}
Rich Branson, Jeffrey~R. Dichter, and et~al. Henry~Feldman.
\newblock {The US Strategic National Stockpile Ventilators in Coronavirus
  Disease 2019: A Comparison of Functionality and Analysis Regarding the
  Emergency Purchase of 200,000 Devices}.
\newblock {\em
  \href{https://journal.chestnet.org/article/S0012-3692(20)34505-0/fulltext}{Chest}},
  2020.
\newblock S0012-3692(20)34505-0.

\bibitem{beitler2020}
J.~R. Beitler, A.~M. Mittel, and et~al. Kallet, R.
\newblock {V}entilator {S}haring during an {A}cute {S}hortage {C}aused by the
  {C}{O}{V}{I}{D}-19 {P}andemic.
\newblock {\em
  \href{https://www.atsjournals.org/doi/ref/10.1164/rccm.202005-1586LE}{Am J
  Respir Crit Care Med}}, 202(4):600--604, 2020.

\bibitem{review_open_ventilators}
J.~M. Pearce.
\newblock {A review of open source ventilators for COVID-19 and future
  pandemics}.
\newblock {\em
  \href{https://doi.org/10.12688/f1000research.22942.2}{F1000Research}}, 9:218,
  2020.

\bibitem{Harris2020}
Matthew Harris, Yasser Bhatti, and et~al. Buckley, Jim.
\newblock {Fast and frugal innovations in response to the COVID-19 pandemic}.
\newblock {\em \href{https://doi.org/10.1038/s41591-020-0889-1}{Nature
  Medicine}}, 26(6):814--817, Jun 2020.

\bibitem{Mora2020}
Simone Mora, F{\'a}bio Duarte, and Carlo Ratti.
\newblock {Can Open Source Hardware Mechanical Ventilator (OSH-MVs) initiatives
  help cope with the COVID-19 health crisis? Taxonomy and state of the art}.
\newblock {\em
  \href{http://www.sciencedirect.com/science/article/pii/S2468067220300596}{HardwareX}},
  8:e00150, 2020.

\bibitem{cern_ventilator}
J.~Buytaert, A.~Abed Abud, and et~al. K.~Akiba.
\newblock {The HEV Ventilator Proposal}, 2020.
\newblock Available:\href{https://arxiv.org/abs/2004.00534}{arXiv:2004.00534}.

\bibitem{milano_ventilator}
C.~Galbiati, W.~Bonivento, and et~al. Caravati, M.
\newblock {Mechanical Ventilator Milano (MVM): A Novel Mechanical Ventilator
  Designed for Mass Scale Production in Response to the COVID-19 Pandemics}.
\newblock {\em
  \href{https://www.medrxiv.org/content/early/2020/03/27/2020.03.24.20042234}{medRxiv:2020.03.24.20042234}},
  2020.

\bibitem{nasa_ventilator}
{California Institute of Technology}.
\newblock {VITAL}: The {COVID-19 V}entilator {D}evice, 2020.
\newblock Available: \url{https://medeng.jpl.nasa.gov/covid-19/ventilator/}.

\bibitem{portugese_ventilator}
Américo Pereira, Luís Lopes, and et~al. Paulo~Fonte.
\newblock {Prototype of an affordable pressure-controlled emergency mechanical
  ventilator for COVID-19}.
\newblock {\em \href{https://arxiv.org/abs/2004.00310}{arXiv:2004.00310}},
  2020.

\bibitem{corovent}
{Czech Technical University, MICo Group}.
\newblock {CoroVent}.
\newblock Available: \url{https://www.micomedical.cz/}.

\bibitem{Raymond2020}
Samuel~J Raymond, Trevor Wesolowski, and et~al. Baker, Sam.
\newblock {A low-cost, rapidly scalable, emergency use ventilator for the
  COVID-19 crisis}.
\newblock {\em
  \href{https://www.medrxiv.org/content/early/2020/09/25/2020.09.23.20199877}{medRxiv:2020.09.23.20199877}},
  2020.

\bibitem{oxvent_ventilator}
{OxVent}.
\newblock Available: \url{https://oxvent.org/product/}.

\bibitem{rice_ventilator}
{Rice University}.
\newblock {ApolloBVM - Emergency Use Ventilator}, 2020.
\newblock Available: \url{http://oedk.rice.edu/apollobvm/}.

\bibitem{open_ventilator}
{The Open Ventilator}, 2020.
\newblock Available: \url{https://www.theopenventilator.com/}.

\bibitem{marburg_ventilator}
Enrique Castro-Camus, Jan Ornik, and et~al. Mach, Cornelius.
\newblock {Simple ventilators for emergency use based on Bag-Valve pressing
  systems: Lessons learned and future steps}.
\newblock {\em
  \href{{https://www.medrxiv.org/content/early/2020/05/14/2020.04.29.20084749}}{medRxiv:2020.04.29.20084749}},
  2020.

\bibitem{MADVent}
Aditya Vasan, Reiley Weekes, and Acute Ventilation Rapid Response~Taskforce
  (AVERT).
\newblock {MADVent: A low-cost ventilator for patients with COVID-19}.
\newblock {\em
  \href{https://onlinelibrary.wiley.com/doi/abs/10.1002/mds3.10106}{MEDICAL
  DEVICES \& SENSORS}}, 3:e10106, 2020.

\bibitem{EVent-2020}
Albert~H. Kwon, Alexander~H. Slocum, and on~behalf of~the MIT E-Vent~Team.
\newblock {Rapidly scalable mechanical ventilator for the COVID-19 pandemic}.
\newblock {\em \href{https://doi.org/10.1007/s00134-020-06113-3}{Intensive Care
  Medicine}}, 46(8):1642--1644, Aug 2020.

\bibitem{nomenclature_draeger}
{Karin Deden}.
\newblock {Ventilation Modes in Intensive Care}.
\newblock
  \href{https://www.draeger.com/Library/Content/rsp-new-nomenclature-ventilation-modes-icu-booklet-9066477-en.pdf}{Dr\"agerwerk
  AG}.

\bibitem{github_HDvent}
{HD}vent github repository.
\newblock Available: \url{https://github.com/HDventilator/}.

\end{thebibliography}
